\documentclass{elsart3}

\usepackage{natbib}

\usepackage{epsfig}

\usepackage{amssymb}

\begin{document}

\begin{frontmatter}



\title{The Gamma--Ray Burst Monitor for Lobster-ISS}

 \author[iasf]{L. Amati\corauthref{cor1}},
 \corauth[cor1]{Corresponding author.}
 \ead{amati@bo.iasf.cnr.it}
 \author[iasf,unife]{F. Frontera},
 \author[iasf]{N. Auricchio},
 \author[iasf]{E. Caroli},
 \author[iasf]{A. Basili},
 \author[sti]{A. Bogliolo},
 \author[unife]{G. Di Domenico},
 \author[iasf]{T. Franceschini},
 \author[unife,liverp]{C. Guidorzi},
 \author[iasf]{G. Landini},
 \author[iasf]{N. Masetti},
 \author[unife]{E. Montanari},
 \author[iasf]{M. Orlandini},
 \author[iasf]{E. Palazzi},
 \author[iasf]{S. Silvestri},
 \author[iasf]{J.B. Stephen},
 \author[iasf]{G. Ventura}
 \address[iasf]{CNR-IASF, Sez. Bologna, via P. Gobetti 101, 40129, Bologna, Italy}
 \address[unife]{Universit\'a di Ferrara, Via Paradiso 12, 44100 Ferrara, Italy}
 \address[sti]{STI, Universit\'a di Urbino, Piazza della Repubblica, 13, 61029 Urbino, Italy}
 \address[liverp]{
Liverpool John Moores University
%Twelve Quays House
Egerton Wharf
Birkenhead CH41 1LD
UK}

\begin{abstract}
% Text of abstract
Lobster-ISS is an X-ray all-sky monitor experiment selected by ESA two years ago
for a Phase A study (now almost completed) for a future flight (2009) aboard the 
Columbus Exposed Payload Facility of the International Space Station. 
The main instrument, based on MCP optics with Lobster-eye geometry, has an 
energy passband from 0.1 to 3.5 keV,  an unprecedented daily sensitivity of 
2$\times$10$^{-12}$ erg cm$^{-2}$ s$^{-1}$, and it is capable to scan, during each orbit,
the entire sky with an angular resolution of 4--6 arcmin.
This X--ray telescope is flanked by a Gamma Ray Burst Monitor, with the minimum 
requirement of recognizing true GRBs from other transient events. In this
paper we describe the GRBM. In addition to the minimum requirement, the instrument proposed
is capable to roughly localize GRBs which occur in the Lobster FOV (162$\times$22.5 degrees)
and to significantly extend the scientific 
capabilities of the main instrument for the study of GRBs and X-ray transients. 
The combination of the two instruments will allow an unprecedented spectral coverage
(from 0.1 up to 300/700 keV) for a sensitive study of the GRB prompt emission 
in the passband where GRBs and X-Ray Flashes emit most of their energy. 
The low-energy spectral band (0.1-10 keV) is of key importance for the study of
the GRB environment and the search of transient absorption and emission features
from GRBs, both goals being crucial for unveiling the GRB phenomenon.
The entire energy band of Lobster-ISS is not covered by either the Swift 
satellite
or other GRB missions foreseen in the next decade.

\end{abstract}


\end{frontmatter}

% main text
\begin{figure*}
\centerline{\epsfig{file=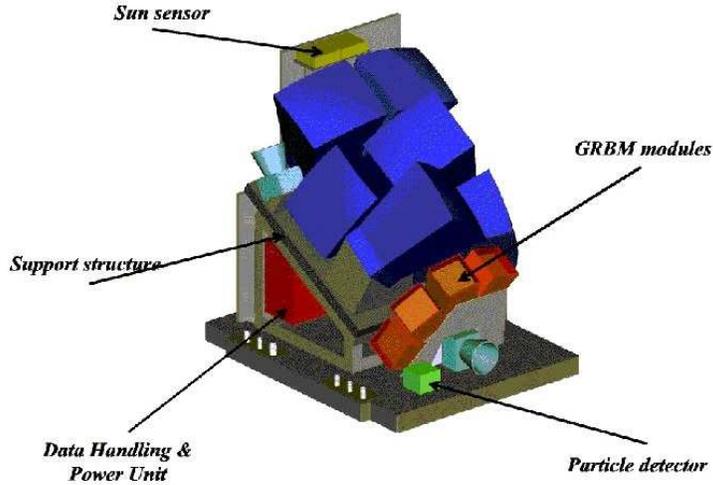,width=10cm}}
\vspace{-2cm}
\caption{Sketch of the Lobster payload. The location of the GRBM  modules is 
indicated (courtesy of Carlo Gavazzi Space).}
\end{figure*}

\section{Lobster--ISS}
\label{lob}
Lobster-ISS \citep{Fraser02} is an X-ray mission proposed by an International 
collaboration led by 
George Fraser (University of Leicester, UK) in response to the ESA call for two 
flexi-missions (F2 and F3) for the International Space Station (ISS). The project 
was approved for an industrial phase A study funded by ESA (kick-off in July 2002,
contractor Carlo Gavazzi Space, Milano) for the accommodation of the experiment 
aboard the ISS Columbus External Payload Facility (CEPF), with its launch in 2009 
and a 3 year duration flight. 
The phase A study, with an additional extension of two months decided by 
ESA in order to allow a better investigation of few 
issues, has been now successfully completed. Unfortunately, given the new policy
by NASA about ISS, the prospects of the Lobster-ISS flight are uncertain.
Because of this, other flight opportunities, like another payload facility aboard ISS or a 
free--flyer satellite, are being investigated.

The main scientific objective of Lobster-ISS is 
the mapping of the X-ray sky in the 0.1--3.5 keV energy band with an angular 
resolution as low as 4--6 arcmin and a daily sensitivity 
of 2$\times$10$^{-12}$ erg cm$^{-2}$ s$^{-1}$ . 
The main instrument is based on Micro-Channel Plate (MCP) optics in a Lobster-eye 
configuration and focal plane detectors based on special sensitive proportional 
counters. It is  composed of six identical modules, each with a Field of View (FOV) 
of 27$^{\circ}$$\times$22.5$^{\circ}$, misaligned in such a way to give a 
total rectangular field of view of 22.5$^{\circ}$ 
in the direction of the ISS motion and 162$^{\circ}$ in the perpendicular direction. 
Thanks to the orbital motion of the ISS, it will be possible to 
map almost all the sky every orbit, allowing the production of a catalog 
of 250000  X-ray sources every two months. The combination of the 
wide field of view, the good angular resolution and the high sensitivity will 
allow the study of the time behaviour of all classes of X-ray sources, 
from comets to stellar coronae, X-ray binaries, 
soft X-ray transients, SNe explosions, AGN, diffuse X-ray background and 
Gamma-Ray 
Bursts (GRBs).
 
Given that many classes of sources (e.g., flare stars, compact X-ray binaries) are
emitters of short transient events, the separation of GRBs from other X-ray short
events is a hard task for an instrument with a passband from 0.1 to 3.5 keV. In order
to overcome this issue, a Gamma Ray Burst Monitor (GRBM) flanks the Lobster-eye
telescope. In this paper we describe the GRBM proposed for the phase A study, its
science goals and performance.

\section{The Lobster-ISS GRBM: scientific goals and design}
\label{sci}

\begin{figure}
\centerline{\epsfig{file=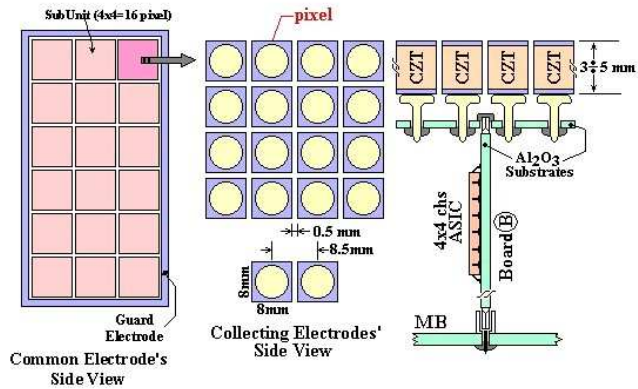,width=9cm}}
\vspace{-3cm}
\caption{
A possible assembly of a CZT-Unit and of a CZT Sub-Unit (16 CZT elementary 
detectors) is shown. The mother board connects CZT sub--units and also houses the 
analog post processing electronics.
}
\end{figure}

The GRBM we propose not only satisfies the minimum requirement of
identifying true GRBs, but
it extends significantly the science objectives of the Lobster-ISS mission.
It consists of 4 misaligned detection units, each one made of an array of 
Peltier-cooled CdZnTe (CZT) detectors, surmounted by a passive collimator which 
defines the field of view. 
The FOV of each unit is 35$^{\circ}$$\times$55$^{\circ}$ 
(FWHM), resulting in a total rectangular field of view of 35$^{\circ}$ in the 
direction of the ISS motion and 240$^{\circ}$ in the perpendicular direction. 
Each  collimator is made of four slabs of 1.5 mm thick Tungsten with a graded
shield to minimize the X--ray fluorescence. To get the above FOV, 
the height of each slab 
is required to be 14.2 cm. 
The use of collimators allows the reduction of the (primary and atmospheric)
background radiation entering through their aperture,  and  reconstruction 
of the burst position within a few degrees (for strong GRBs)
from the different direction of the collimator axes. 
With this instrument configuration, a source is viewed by at least 
2 detection units for most 
(95\%) of the directions within the FOV of Lobster. As a consequence, the area 
exposed by the GRBM to these directions ranges from 0.70 to 1.30 times the useful 
area of a single detection unit.

\begin{center}
\begin{table}
\small
\caption{GRBM scientifically-relevant specifications}
\begin{tabular}{ll}
\hline
No. of  units & 4 \\
Detection area/unit & 184 cm$^2$ \\
Energy band & 3--300/700 keV \\
Instantaneous field of view & 35$^{\circ}$ x 55$^{\circ}$ (FWHM) 
\\
 & (per module) \\
 & 35$^{\circ}$ x 240$^{\circ}$ for 4 modules \\
Total exposed area to a given point & from 129 to 240 cm$^2$  \\
Detector & Cooled  ($\sim$250 K) CZT  \\
Energy resolution & $\Delta$E/E$\sim$3\% @ 6 keV \\
Minimum (non-zero) exposure time & $\sim$1150 s \\
for a given point, per orbit & \\
\hline
\end{tabular}
%\vspace{0.2cm}
\end{table}
\end{center}

A sketch of the possible Lobster--ISS payload
configuration is shown in Figure~1, where the blue and red modules correspond
to the X--ray telescope and the GRBM, respectively.
Each detection unit (see Figure~2) is made of an array of CZT elementary 
crystals (pixels) \citep{Caroli99}. Each pixel has a cross section  of 8$\times$8 mm$^2$. 
The elementary crystals 
are packed together in sub-units of 4$\times$4 pixels, while each detection unit is 
made of 3$\times$6 sub-units. Thus the X-ray sensitive array is made of 288 pixels. The 
active area of each unit is  184 cm$^2$  while its geometric area, which takes into 
account a 0.5 mm pitch between each couple of pixels, is 208 cm$^2$. The crystals are 
assembled on thin (1 mm) ceramic plates. Below each module is located the 
front-end electronics with multiplexers and ADCs. 
The detector thickness, which was assumed for the phase A study in order to guarantee
an instrument passband from 3 to 300 keV, is 5~mm. We are evaluating the
extension of the energy passband up to 700 keV either by increasing the CZT thickness
up to 10 mm or by using an alternative detector. The very valid alternative to
the CZT is a phoswich-like detector made of a silicon drift chamber coupled 
with a CsI scintillator, the development of which is underway \citep{Marisaldi04}.
The scientifically relevant specifications of the baseline configuration of the 
GRBM are summarized in Table~1.
In the case of GRB detection, the GRBM electronics and on-board data handling provide
high resolution spectra and light curves. The GRB position is also determined on board
and automatically 
transmitted to the Lobster data handling electronics for a better determination 
by means of the X-ray telescope. 

Thanks to the GRBM, the Lobster-ISS passband extends from 0.1 to 300/700 keV,
an unprecedented energy band never used by an all-sky monitor 
to study  the prompt emission of GRBs. The very good energy resolution of
the GRBM, 3\% at 6 keV,
will permit a sensitive study  of transient 
absorption and/or emission features during the early phase of the prompt emission.
Moreover, the combination of the two instruments will allow an unprecedented study
of the absorption cut--offs below 2 keV in the GRB spectra.
As demonstrated by BeppoSAX \citep{Amati00,Frontera01,Frontera04}, the study of these 
features and cut-offs is of key 
importance for the determination of the circumburst environment properties, the nature of 
progenitors and the connection with SNe. The proposed GRBM, given its multi-pixel
configuration, can also be exploited to measure the polarization of the GRB prompt 
emission 
by optimizing the CZT thickness for this goal \citep{Curado04}. 
In the described configuration, 
Lobster-ISS is also the ideal mission for studying X-Ray Flashes (XRFs) and their 
nature \citep{Kippen01,Barraud03}. In addition to GRBs and XRFs, the 
GRBM will allow identification and study of 
all types of fast high energy X--ray transients, in particular of the
Soft Gamma Repeaters \citep{Feroci99,Guidorzi04}.

\section{Expected performances}
\label{per}

\begin{table*}
\caption{Expected background level and 5$\sigma$ sensitivity to GRB events, computed by 
assuming a typical GRB spectrum.}
\begin{center}
\begin{tabular}{cccc}
\hline
Energy band & Background & Flux sensitivity & Flux sensitivity \\
(keV) & (cts cm$^{-2}$ s$^{-1}$) & (photons cm$^{-2}$ s$^{-1}$) & (10$^{-8}$ erg cm$^{-2}$ s$^{-1}$) \\
\hline
3--10 & 6.7 & 2.0 & 1.8 \\
10--30 & 2.1 & 1.2 & 3.5 \\
30--200 & 1.0 & 0.85 & 10 \\
50--300 & 0.6 & 0.68 & 13 \\
\hline
\end{tabular}
\end{center}
\end{table*}

\subsection{Background and sensitivity}

The mean background level which is expected at the ISS orbit for the instrument 
configuration described above is reported in Table~2. It has been estimated by 
summing the contributions of the
diffuse X--ray background and the intrinsic (particle--induced) background.
The intrinsic background spectrum is assumed to 
have a power law shape with count 
index of $-$1.4 (value based on past experience) and normalization derived 
from the assumption that the count intensity 
in the 30--200 keV range is 3$\times$10$^{-3}$ cts cm$^{-2}$ s$^{-1}$ .
Given the wide field of view of the GRBM, the contribution of bright galactic sources 
to the total background level is not negligible. In particular, when the Galactic 
Bulge is inside the FOV of a detection unit, the background level is expected to 
increase by about 30\%. This has been taken into account in the estimates and 
simulations reported below. \\
%\subsection{Flux Sensitivity}
The 5$\sigma$ sensitivity to GRB events is reported in Table~2. Given that a 
GRB in the FOV of the Lobster telescope is viewed by at least 2 GRBM detection 
units, we assumed the background level corresponding to an area of 2 x 184 = 368 
cm$^2$ . 
Conservatively, we assumed for the area exposed to the source the minimum value
of 129 cm$^2$, 
corresponding to 70\% of the area of each detection unit (see Sect. 2). 
As source spectrum we assumed a typical Band function \citep{Band93} with 
$\alpha$= $-$1, $\beta$ = $-$2 and E$_0$ = 200 keV.
The 5$\sigma$ sensitivity of the GRBM to a Crab-like source as a function of 
exposure, again assuming a total illuminated area of 129 cm$^2$, is shown in 
Figure~3
for the cases
of a source located in the galactic center region and of a source with high galactic 
latitude.
We note that, by means of an off--line refined analysis, we expect to be able to 
identify 
the pixels (or groups of pixels) illuminated by the source. In this case the 
detection area contributing to the background level will be reduced to that exposed 
to the source and the instrument sensitivity will be significantly improved with 
respect to the
values reported in Table~2 and Figure~3.

\begin{figure}
\centerline{\epsfig{file=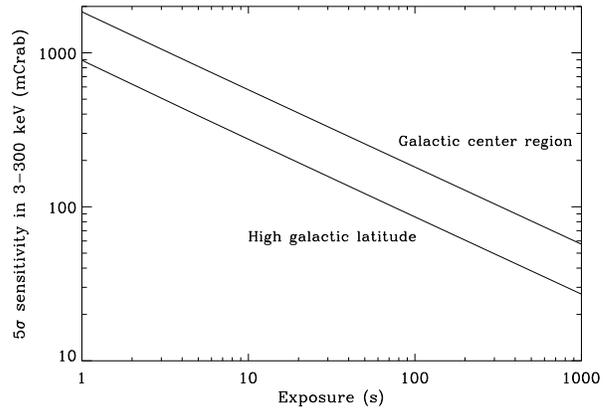,width=9cm}}
\vspace{-6.0cm}
\caption{
Expected 5$\sigma$ sensitivity of the GRBM  to a Crab--like source as a 
function of exposure time. Two  cases have been considered: a source located in the 
galactic center region and source at high galactic latitude.
}
\end{figure}

\subsection{Source localization and discrimination}

The simulated source localization accuracy (90\% confidence level, c.l.) as a function 
of the burst fluence in the 50--300 keV energy band ranges 
from $\sim$1--2$^{\circ}$ 
for the brightest events to several tens of degrees for the weakest ones.
In these simulations we assumed the values of the total detector 
area illuminated by the source, background level and source spectrum  assumed for the
flux sensitivity evaluation. We used the source localization reconstruction algorithm 
expected to be used during the flight and based on that adopted for the
BeppoSAX/GRBM \citep{Guidorzi01}.

As mentioned in Section 2, the minimum requirement for the GRBM is its capability of 
identifying true GRBs. To this end, exploiting the BeppoSAX 
experience, 
we intend to set a constraint to the hardness ratio (HR) between the count rates in two energy 
bands  \citep{Guidorzi01}.  By means of numerical simulations, we find that the 
significance of the ratio between the expected count rates in the 30-70 keV
and  70-200 keV energy bands is sufficient to allow the  discrimination between GRBs and 
other transient events, in the range of fluences and durations detectable with the Lobster 
telescope in 1 s.

\begin{figure}
\centerline{\epsfig{file=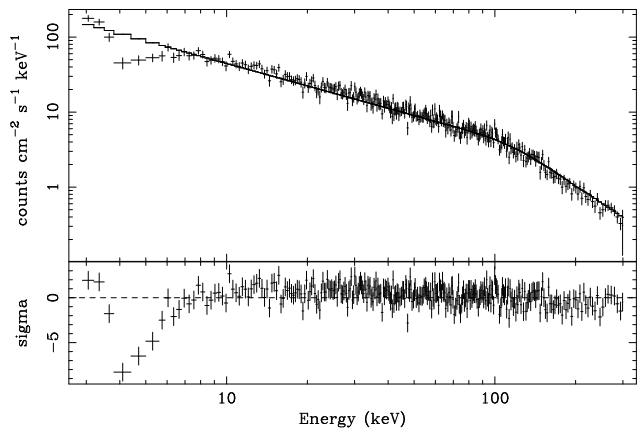,width=7cm,angle=-90}}
\caption{
The spectrum of GRB 990705 in the first 13 s \citep{Amati00}
as expected  to be measured with the  proposed GRBM for Lobster-ISS. The absorption 
edge at $\sim$3.8 keV is apparent at a  $>$ 12$\sigma$ significance.
}
\end{figure}

\subsection{Sensitivity to spectral features}

One of the main goals of the present GRBM configuration is the possibility of 
detecting transient absorption features during the rise time of the GRB events. 
Such features, possibly associated with a variable column density, are predicted by
several models and have already been observed in two GRBs \citep{Amati00,Frontera04}. 
The joint fitting of the GRBM and Lobster 
spectra will extend the  analysis down to 0.1 keV, allowing not only the 
study of the column density behaviour with time, but also the increase of the 
edge significance. In Figure~4 we show the spectrum  
expected to be detected by the GRBM for Lobster-ISS using as template that measured, 
with the BeppoSAX WFC plus GRBM, from GRB~990705 during the first 13 s. The presence 
of the feature is visible at much higher statistical significance than that observed 
with BeppoSAX \citep{Amati00}. 
 \\
%\subsection{Sensitvity to emission features in the GRB spectra}
In addition to absorption features, also transient emission components are expected 
to contribute to the early GRB emission, in particular black--body emission from the 
photosphere of the fireball. Evidence of such a transient  
broad emission component was actually found in the prompt emission  of 
GRB~990712, and could be modeled  with a blackbody with temperature kT$\sim$1.3 keV 
\citep{Frontera01}. We simulated a GRBM spectrum by assuming 
as template the spectrum of GRB~990712 in the time interval in which the transient 
emission feature was observed. The excess with respect to the power-law continuum 
is detected at $\sim$8$\sigma$ significance. As above, for these simulations 
we assumed a total exposed area of 129 cm$^2$ (worst case).

\subsection{Sensitivity to polarization}

The polarization measurement of the prompt gamma-ray emission from GRBs is
recognized to be one of the major objectives of GRB studies, being of key
importance to establish the emission mechanism of the radiation (e.g. synchrotron,
inverse Compton). 
For this purpose, it is
important to measure: i) the spectral dependence of the linear polarized
fraction and ii) the temporal dependence of the position angle. These
quantities would allow to clearly identify the emission mechanism, the
geometry of the advected magnetic field and the origin of the light curve
variability. In addition, it is clearly of primary importance to measure
the level of polarization in the GRBs for which the observed spectrum is
supposed to be in violation of the synchrotron properties. 
With the present configuration, for  a GRB with 25-100 keV fluence 
similar to that of GRB~021206 (2.9$\times$10$^{-5}$ erg cm$^{-2}$, Coburn 
\& Boggs, 2003), the expected minimum detectable linear polarization 
is  $\sim$60\% in the 70--150 keV 
energy range and $\sim$30\% in the 150--300 keV band.

\end{document}